\documentclass[11pt,preprintnumbers,nofootinbib,prd]{revtex4-1}
\usepackage{color}
\usepackage{graphicx}
\usepackage{amsmath}
\usepackage{epsfig}
\usepackage{mathrsfs}
\usepackage{float}

\newcommand{\be}{\begin{equation}}
\newcommand{\ee}{\end{equation}}
\newcommand{\ba}{\begin{eqnarray}}
\newcommand{\ea}{\end{eqnarray}}

\usepackage{hyperref}
\usepackage{rotating}
\usepackage[all]{xy}
\usepackage{amssymb,euscript,bbold}
\newcommand{\CP}{{\ensuremath{\mathbb{P}}}}
\newcommand{\Xt}{{\ensuremath{\widetilde{X}}}}
\newcommand{\Dt}{{\ensuremath{\widetilde{D}}}}
\newcommand{\C}{\ensuremath{\mathbb{C}}}
\newcommand{\R}{\ensuremath{\mathbb{R}}}
\newcommand{\Z}{\ensuremath{\mathbb{Z}}}
\newcommand{\Osheaf}{\ensuremath{\mathscr{O}}}
\DeclareMathOperator{\Span}{span}
\DeclareMathOperator{\Kc}{Kc}

\DeclareMathOperator{\ch}{ch}
\DeclareMathOperator{\Td}{Td}
\DeclareMathOperator{\diag}{diag}
\DeclareMathOperator{\Vol}{Vol}
\DeclareMathOperator{\re}{Re}
\DeclareMathOperator{\im}{Im}

\newcommand{\hplus}[1]{\ensuremath{h^{#1}_{+}}}
\newcommand{\hminus}[1]{\ensuremath{h^{#1}_{-}}}
\newcommand{\Tbar}{\ensuremath{\overline{T}}}

\newcommand{\MKahler}{\ensuremath{{\cal M}_K}}
\newcommand{\TeV}{\ensuremath{\,\text{TeV}}}
\newcommand{\GeV}{\ensuremath{\,\text{GeV}}}

\newcommand{\asusy}[1]{\ensuremath{a_{#1}^\text{susy}}}
\newcommand{\amin}[1]{\ensuremath{a_{#1}^\text{min}}}
\newcommand{\tsusy}[1]{\ensuremath{t_{#1}^\text{susy}}}

\newcommand{\taususy}[1]{\ensuremath{\tau_{#1}^\text{susy}}}
\newcommand{\taumin}[1]{\ensuremath{\tau_{#1}^\text{min}}}
\newcommand{\chisusy}[1]{\ensuremath{\chi_{#1}^\text{susy}}}
\newcommand{\chimin}[1]{\ensuremath{\chi_{#1}^\text{min}}}

\newcommand{\eqdef}{%
 \mathrel{\lower.1mm
   \hbox{$\stackrel{\lower.424ex\hbox{\scriptsize def}}{=}$}}
}

\begin{document}
\begin{titlepage}
  \vspace*{-2cm}
  \hfill
  \parbox[c]{5cm}{
    \begin{flushright}
      OHSTPY-HEP-T-10-004\\
      UCB-PTH-10/06\\
      DIAS-STP 10-02
    \end{flushright}
  }
  \vspace{-1cm}
  \vspace*{\stretch1}
  \begin{center}
    \Huge
    Stabilizing All K\"ahler Moduli\\
    in Type IIB Orientifolds
  \end{center}
  \vspace*{\stretch1}
  \begin{center}
    \begin{minipage}{\textwidth}
      \begin{center}
        \sc
        Konstantin Bobkov$^1$,
        Volker Braun$^2$,\\
        Piyush Kumar$^3$, and
        Stuart Raby$^1$
        \\[5ex]
        \it
        $^1$Department of Physics\\
        The Ohio State University\\
        Columbus, OH 43210, USA
        \\[3ex]
        $^2$Dublin Institute for Advanced Studies\\
        10 Burlington Road\\
        Dublin, Ireland
        \\[3ex]
        $^3$Department of Physics\\
        University of California\\
        $\&$\\
        Theoretical Physics Group\\
        Lawrence Berkeley National Laboratory\\
        Berkeley, CA 94720 USA
      \end{center}
    \end{minipage}
  \end{center}
  \vspace*{\stretch1}
  \begin{center}
    \textbf{Abstract}
    \\[2ex]
    \begin{minipage}{0.75\textwidth}
      \setlength{\parindent}{3ex} We describe a simple and robust
      mechanism that stabilizes \emph{all} K\"ahler moduli in
      Type IIB orientifold compactifications. This is shown to be
      possible with just one non-perturbative contribution to the
      superpotential coming from either a D3-instanton or D7-branes
      wrapped on an ample divisor. This moduli-stabilization mechanism 
      is similar to and motivated by the one used in the fluxless
      $G_2$ compactifications of $M$-theory. After explaining the general idea,
      explicit examples of Calabi-Yau orientifolds with one and three
      K\"ahler moduli are worked out. We find that the stabilized
      volumes of \emph{all} two- and four-cycles as well as the volume
      of the Calabi-Yau manifold are controlled by a single parameter,
      namely, the volume of the ample divisor. This feature would
      dramatically constrain any realistic models of particle physics
      embedded into such compactifications.  Broad consequences for
      phenomenology are discussed, in particular the dynamical
      solution to the strong CP-problem within the framework.
    \end{minipage}
  \end{center}
  \vspace*{\stretch1}
\end{titlepage}
\tableofcontents

\section{Introduction}
\label{intro}

One of the central goals of string phenomenology is to explain how
particular values of low-energy physics parameters, such as the
fine-structure constant or the electron Yukawa coupling, arise from a
fundamental theory with no free parameters. Within the context of
string theory, the values of these parameters are tied to the vacuum
expectation values (vevs) of \emph{moduli}. However, at a classical
level\footnote{with no background fluxes}, the moduli fields are
massless and do not have fixed non-zero vevs. 
Therefore, in order to be able to explain the values of low-energy 
physics parameters and to be able to do realistic phenomenology, 
the moduli must be stabilized. The issue of moduli stabilization 
also has important implications for supersymmetry breaking and 
the cosmological constant.

Considerable progress has been made in the field of moduli
stabilization within various corners of string theory, such as Type
IIA~\cite{DeWolfe:2005uu, Villadoro:2005cu, Camara:2005dc,
  Acharya:2006ne, Silverstein:2007ac}, Type IIB~\cite{Kachru:2002he,
  Kachru:2003aw, Denef:2004dm, Denef:2005mm, Balasubramanian:2005zx,
  Conlon:2005ki, Lust:2005dy, Lust:2006zg},
Heterotic~\cite{Gukov:2003cy, Becker:2004gw, deCarlos:2005kh,
  Curio:2005ew, Micu:2009ci} and $G_2$ compactifications of
$M$-Theory~\cite{Acharya:2002kv, Acharya:2008hi, Acharya:2007rc,
  Acharya:2006ia, deCarlos:2004ci}. The simplest recipe for moduli
stabilization and constructing vacua with a small positive
cosmological constant (de Sitter vacua) within Type IIB string theory
was proposed by Kachru, Kallosh, Linde, and
Trivedi~\cite{Kachru:2003aw} (KKLT). They considered a toy model with
one K\"ahler modulus in which the dilaton and complex structure moduli
were stabilized by flux contributions to the
superpotential~\cite{Giddings:2001yu, Dasgupta:1999ss}, while the
K\"ahler modulus $T$ was stabilized by a non-perturbative contribution
to the superpotential. The non-perturbative contribution can in
general arise from Euclidean D3-brane instantons or from strong gauge
dynamics on stacks of D7-branes wrapping four-cycles, or divisors
$D_i$, inside the Calabi-Yau manifold, when certain topological
conditions are satisfied.

It is commonly believed, at least in the simplest setup where the
K\"ahler potential contains no perturbative $\alpha^{\prime}$ or
string loop corrections, that it is necessary for the non-perturbative
part of the superpotential to contain at least $\hplus{11}$ linearly
independent divisors $D_i$ in order to fix all $\hplus{11}$ K\"ahler
moduli $\tau_i=\re T_i$. Thus, one imagines the following
superpotential:
\begin{equation}\label{eq:W0plusInst1}
  W = W_0 + 
  \sum_i A_ie^{-\frac{2\pi}{c_i}\sum_k n^{(i)}_k T_k }
  ,
\end{equation}
where $W_0, A_i, c_i, n^{(i)}_k$ are constants (more on this in
\autoref{sec:single}). Following the KKLT proposal, several explicit
examples of Calabi-Yau manifolds with few K\"ahler moduli were
constructed in which the above approach to moduli fixing was
successfully implemented~\cite{Denef:2004dm, Denef:2005mm}.  A
consequence of the above mechanism is that the pseudoscalar partners
of the K\"ahler moduli $\chi_i = \im T_i$ are also generically
stabilized with masses comparable to that of the $\tau_i$.

Although the above mechanism works for simple cases with a few
K\"ahler moduli, one faces a number of challenges in extending it to
quasi-realistic string compactifications, which could describe
low-energy particle physics. First, realistic compactifications
describing the many low energy parameters of particle physics are
expected to contain many moduli.  Thus, one generically expects the
Calabi-Yau manifold $X$ to contain many K\"ahler moduli, with
$\hplus{11}(X) ={\cal O}(100)$. Finding explicit examples with such a
large number of linearly independent divisors contributing to the
superpotential is a daunting task, as one must ensure that the
appropriate topological condition (zero-mode structure) is satisfied
for each linearly independent divisor. Second, as pointed out
in~\cite{Blumenhagen:2007sm}, the K\"ahler modulus, which measures the
volume of any four-cycle containing a chiral spectrum \emph{may not}
be stabilized purely by non-perturbative effects in the superpotential
as in eq.~\eqref{eq:W0plusInst1}. This is particularly relevant for
the visible sector four-cycle as it contains a chiral spectrum. In
such cases, the correct counting of zero-modes implies that
exponential terms in the above superpotential appear with
field-dependent prefactors containing gauge invariant combinations of
chiral matter fields.  Thus, in order to fix all such four-cycles, the
matter fields must also be dynamically fixed in a phenomenologically
viable way. However, for the visible sector such operators must have
vanishing vevs for phenomenological reasons; therefore the
corresponding superpotential contributions are zero.  Finally, even if
one comes up with a mechanism to stabilize such K\"ahler moduli, say
by perturbative $\alpha'$ or string-loop effects~\cite{Cicoli:2008va},
a detailed analysis of vacua generated by such a superpotential
depends on a large number $\gtrsim \hplus{11}$ of independent parameters
that enter the superpotential, making it quite intractable to make
robust predictions relevant for particle phenomenology.
 
In this work, we advocate a different approach to fixing the K\"ahler
moduli that is largely motivated by the general results obtained in
the fluxless $G_2$ compactifications of M-theory~\cite{Acharya:2008hi}.
There, it is possible to stabilize {\it all} the moduli even when
the non-perturbative superpotential receives contributions from a 
{\it single} associative three-cycle in a form of two
gaugino condensates, as long as the three-cycle intersects {\it all} 
co-associative four-cycles positively.
Thus, here we consider Calabi-Yau orientifolds $X$ containing fewer than
$\hplus{11}(X)$ divisors contributing to the superpotential. In the
extreme case it is possible to restrict to a single divisor $D\in X$,
which contributes to the non-perturbative superpotential in a form of
a gaugino condensate or an instanton. The main claim of the paper is
that even though the superpotential depends only on a single linear
combination of K\"ahler moduli $\tau_i$, they can be stabilized
self-consistently while satisfying the supergravity approximation. To
be precise, we will show that this is the case if and only if the
divisor $D$ is \emph{ample}. Then all the four-cycle volumes $\tau_i$
and two-cycle volumes $t_i$ will be automatically stabilized inside
the K\"ahler cone. We will define and explain this in detail in
\autoref{super}.

The plan of the paper is as follows. The contributions to the
superpotential and the conditions required are reviewed in
\autoref{super} within the context of the original KKLT
proposal. \autoref{coords} discusses some general properties of the
K\"ahler moduli space of Type IIB orientifolds which are crucial in
performing explicit computations without specifying a particular
orientifold. In \autoref{minimize}, the minimization of the scalar
potential and stabilization of moduli within the above framework is
described in detail, where the results of \autoref{coords} are
utilized. In \autoref{sec:example}, explicit examples of Calabi-Yau
orientifolds realizing the moduli stabilization procedure described in
\autoref{sec:single} are provided. The general formalism developed in
\autoref{sec:single} is applied to these particular cases. In
\autoref{consistency}, we describe conditions under which our results 
hold parametrically even in general cases with multiple contributions
to the superpotential. Finally, \autoref{phenocosmo} is a brief
discussion of the broad phenomenological consequences of this
framework, followed by conclusions in \autoref{conclude}.

\section{Stabilizing All K\"ahler Moduli with a Single Non-Perturbative Contribution}
\label{sec:single}

\subsection{Contributions to the Superpotential}
\label{super}

In the proposal of KKLT, the superpotential is given by a sum of the
tree-level flux contributions~\cite{Gukov:1999ya, Taylor:1999ii,
  Curio:2000sc} that fix the dilaton and complex structure
moduli~\cite{Giddings:2001yu, Dasgupta:1999ss},
\begin{equation}
  W_0 = \int_X G_3\wedge\Omega
  ,
\end{equation}
combined with non-perturbative contributions generated by strong gauge
dynamics on stacks of D7-branes or Euclidean D3-brane instantons wrapping
divisors $D_i\subset X$. Their volumes are
\begin{equation}
  \Vol(D_i) = \re \left(\sum_k n^{(i)}_k T_k\right)
  ,
\end{equation}
where $n^{(i)}_k \in \Z$ specify the homology class $[D_i]\in H_4(X,\Z)$
of the $i$-th divisor. The combined superpotential is then given by
\begin{equation}
  \label{eq:W0plusInst}
  W = W_0 + 
  \sum_i A_ie^{-\frac{2\pi}{c_i}\sum_k n^{(i)}_k T_k }
  ,
\end{equation}
where $c_i$ is the dual Coxeter number of the condensing gauge
group\footnote{For example, $c_i=N$ for $SU(N)$ or $c_i=1$ for a
  single instanton.}. The complex-structure and dilaton moduli are
stabilized at a high scale, close to the Kaluza-Klein scale. Moreover,
although $W_0$ is generically ${\cal O}(1)$ in string units, it can be
shown that by choosing special values of the integer-quantized fluxes
it is possible to tune $W_0$ to a small value~\cite{Giryavets:2003vd, Giryavets:2004zr}. 
This turns out to be crucial, as will be explained later.

Moving on to the non-perturbative contribution, it is important to
note that not every D3-instanton (or gaugino condensate on a
spacetime-filling D7 brane) wrapped on a divisor $D\subset X$
contributes to the superpotential. One has to analyze the zero-modes,
which is more complicated than the analogous count for
M5-branes~\cite{Witten:1996bn}. In particular, it depends not only on
the arithmetic genus $\chi(D,\Osheaf_D)$ but also on the details of
the orientifold action~\cite{Bergshoeff:2005yp, Park:2005hj,
  Lust:2006zg}. Note that one can use the orientifold to split the
arithmetic genus into
\begin{equation}
  \chi_\pm(D,\Osheaf_D) = 
  \sum_{n=0}^2 (-1)^n h^n_\pm\big(D,\Osheaf_D\big)
  .
\end{equation}
In the absence of fluxes or intersections with other D-branes, a
\emph{necessary} condition for a D3-instanton in an O3/O7 orientifold
to contribute is~\cite{Blumenhagen:2010at, Blumenhagen:2010ja}
\begin{equation}
  \label{necessary} 
  \chi_+-\chi_-=1.
\end{equation} 
Within the above setup, background fluxes are required to be present
in order to stabilize the complex structure and dilaton moduli and
give rise to the term $W_0$ in eq.~\eqref{eq:W0plusInst}. However, the
above criterion still holds if the fluxes are such that they do not
change the zero mode counting of the D3-instanton. For simplicity,
this will be assumed from now on. Moreover, the coefficient $A$ might
be zero depending on the topology of the instanton moduli space. For
simplicity, we will assume that $A\not=0$ and is an $O(1)$ complex
number in the remainder of this paper.

In fact, ``fractional instantons'' can relax the above condition on
the divisor $D$. Their microscopic origin is a stack of D7-branes,
which fill spacetime in addition to wrapping the divisor, and give
rise to gaugino condensation at low energies. In this case, a divisor
with $\chi_+-\chi_- \geq 1$ might also contribute to the
non-perturbative superpotential under some
circumstances~\cite{Gorlich:2004qm}. In particular, gaugino
condensation can still occur in the presence of charged matter on the
stack if the charged matter gets a mass at sufficiently high scales
leaving a pure gauge theory at low energies. In fact, this is to be
expected if the matter is vector-like. The effect on the
superpotential will be an exponential term with $c_i>1$ in
eq.~\eqref{eq:W0plusInst}. Thus if $\chi_+-\chi_-=1$ for a given
divisor, there is either a D3-instanton contribution (in the absence
of further branes) or a gaugino condensate (if there are stacks of
D7-branes wrapping the divisor). On the other hand, for
$\chi_+-\chi_->1$ the only non-perturbative contribution could arise
from a stack of D7-branes (with an appropriate spectrum) wrapping the
divisor. The formal computation of the moduli vevs is the same in both
cases, however, since we can effectively rescale $T_k \mapsto c_i T_k$
for the case of a gaugino condensate. For concreteness, therefore, we
will mostly discuss D3-instantons.

A \emph{sufficient} (but overly strong) condition\footnote{Here and in
  the following we will implicitly always assume that $D$ is smooth
  and maps to itself under the orientifold action.  Moreover, the
  orientifold planes are only O3/O7 (no O5/O9).} for the D3-instanton
to contribute to the superpotential is to demand that it is rigid and
ample. Let us quickly review these notions:
\begin{itemize}
\item A divisor $D$ is \emph{rigid} if it cannot be deformed. Thinking
  of the divisor as the zeroes of a section in a line bundle
  $\Osheaf(D)$, this is precisely the case if the section is unique up
  to an overall constant, that is, $h^0\big(X,\Osheaf(D)\big)=1$. On a
  Calabi-Yau manifold, this is equivalent to $h^{02}(D)=0$.
\item A divisor is \emph{ample} if and only if the Poincar\'e
  isomorphism $H_4(X,\Z)\simeq H^2(X,\Z)$ identifies it with a
  cohomology class that can be represented by the K\"ahler form of a
  smooth Calabi-Yau metric. We will have much more to say about this
  condition on the next page.

  For now, note that ample divisors enjoy many favorable
  properties. In particular, the Kodaira vanishing forces
  $h^q\big(X,\Osheaf(D)\big)=0$ for all $q>0$, and the Lefshetz
  hyperplane theorem identifies $h^{pq}(X)=h^{pq}(D)$ for all $p+q<2$
  as well as $\pi_i(X)=\pi_i(D)$ for $i\leq 1$.
\end{itemize}
This then leads to the following zero mode spectrum on the divisor,
see eq.~\eqref{eq:hodgeD}:
\begin{itemize}
\item $D$ ample $\Rightarrow$ $h^{00}(D) = 1$ and $h^{01}(D) = 0$.
\item $D$ rigid $\Rightarrow$ $h^{02}(D) = 0$.
\end{itemize}
In particular we see that $\chi_+=1$, $\chi_-=0$, and
eq.~\eqref{necessary} is automatically satisfied.

For simple cases of Calabi-Yau threefolds $X$ with one (or few)
K\"ahler moduli, one may try to construct $\hplus{11}(X)$ rigid and
linearly independent divisors. The corresponding D3-instantons would
contribute $\hplus{11}(X)$ non-perturbative terms in the
superpotential and hence stabilize all the K\"ahler moduli. However,
as pointed out in the introduction it is clear that for realistic
compactifications with large $\hplus{11}(X)$, such a procedure is
quite difficult. The intuition that one needs $\hplus{11}(X)$
linearly-independent divisors $D_i$ contributing to the superpotential
in order to stabilize all K\"ahler moduli comes from Type IIB
compactifications which admit an $F$-theory lift. Within $F$-theory
compactifications on a smooth elliptically fibered Calabi-Yau fourfold
with a Fano base, this is, in fact, a theorem if one only allows
\emph{smooth} divisors contributing to the
superpotential~\cite{Robbins:2004hx}.

Motivated by the general results obtained in fluxless $G_2$
compactifications of M-theory~\cite{Acharya:2008hi}, in this paper we
show that it is possible to stabilize all $\hplus{11}(X)$ K\"ahler
moduli in Type IIB orientifold compactifications containing fewer than
$\hplus{11}(X)$ divisors contributing to the superpotential. In fact,
it is possible to restrict to a single divisor $D\in X$, which
contributes to the non-perturbative superpotential in the form of a
single D3-instanton or gaugino condensate. We will demonstrate that,
contrary to the naive expectation, even when the superpotential
\begin{equation}
  \label{eq:super}
  W\big(T_i\big) = 
  W_0 +
  A e^{-\frac {2\pi}N \sum_{i=1}^{\hplus{11}} n_i T_i }
\end{equation}
depends only upon the single linear combination $\vec{n}\cdot
\vec{T}$, all four-cycle volumes (K\"ahler moduli) $\tau_i$ and all
two-cycle volumes $t_i$ can be stabilized self-consistently while
satisfying the supergravity approximation. For explicitness, we will
assume from now on that there is, indeed, only a single contribution
to the superpotential.

The condition that all volumes of (dimension $2$, $4$, and $6$)
holomorphic submanifolds are positive can be easily formulated in
terms of the K\"ahler form. Its cohomology class $[\omega]\in H^{1,1}(X)$
is parametrized by the K\"ahler moduli, but clearly not all
values\footnote{For example, if $[\omega]$ is a K\"ahler class then
  $-[\omega]$ cannot be the cohomology class of a K\"ahler form.}  are
allowed. The values that can be realized by smooth Calabi-Yau metrics
form the so-called K\"ahler cone
\begin{equation}
  \Big\{
  [\omega]
  ~\Big|~
  \omega = g_{i\bar j}\mathrm{d}z^i \wedge \mathrm{d}z^{\bar j}
  \Big\}
  =
  \Kc(X) 
  \subset H^{1,1}(X)
  .
\end{equation}
In fact, the K\"ahler cone is a convex cone given by a set of linear
inequalities in the 2-cycle volumes $t_i$. In simple cases, like the
examples we will encounter later on, the K\"ahler cone is just the
cone over a simplex. Then, perhaps after a linear change of
coordinates, the K\"ahler cone is
\begin{equation}
  \label{eq:Kc}
  \Kc(X) = 
  \big\{ \vec{t} ~\big|~ t_i >0 \big\}
  \subset H^{1,1}(X)
  .
\end{equation}
We can now be more explicit in the definition of an ample divisor. As
we mentioned above, they are identified with potential K\"ahler
forms. Explicitly, a divisor defines a line bundle whose first Chern
class is the associated cohomology class. Therefore, a divisor $D$ is
ample if and only if $c_1\big(\Osheaf(D)\big) \in \Kc(X)$. If we take
the basis $[D_i]$ of $H_4(X,\Z)$ to be Poincar\'e dual to the
two-cycle classes used in eq.~\eqref{eq:Kc}, then
\begin{equation}
  D=\sum_{i=1}^{\hplus{11}} n_i D_i 
  \quad\text{is ample}
  \qquad \Leftrightarrow \qquad
  n_i>0,
  \quad i=1,\dots,\hplus{11}
  .
\end{equation}
Note that, at this point, we have two seemingly independent elements
of the K\"ahler cone:
\begin{itemize}
\item The (smooth) Calabi-Yau metric defines a class
  $[\omega]\in\Kc(X)\subset H^{1,1}(X)$.
\item The ample divisor $D$ wrapped by the D3-instanton defines a
  class $c_1\big(\Osheaf(D)\big) \in \Kc(X) \subset H^{1,1}(X)$.
\end{itemize}
The point of this paper is to show that the supergravity action, with
the superpotential term generated by $D$, dynamically adjusts the
K\"ahler moduli (that is, the coordinates of $[\omega]$) such that
these are the same up to an overall rescaling of the volume. That is,
the superpotential contribution of a single divisor stabilizes all
K\"ahler moduli at a specific point
\begin{equation}
  [\omega] = \lambda ~\cdot~ c_1\big(\Osheaf(D)\big)
  ,\qquad
  0<\lambda \in \R
  .
\end{equation}
Therefore, the Calabi-Yau metric is smooth (and all volumes are
positive) if and only if $D$ is ample.

Some important comments are warranted. Within the context of the
theorem in $F$-theory mentioned above this implies that these Type IIB
compactifications do not admit a simple lift to $F$-theory with the
stated properties. In particular, the arithmetic genus
$\chi(D,\Osheaf_D)<0$ for a smooth ample divisor in a Calabi-Yau
fourfold~\cite{Robbins:2004hx}. This implies that the $F$-theory lift
of these single-instanton compactifications, if possible, must employ
non-smooth divisors in the corresponding Calabi-Yau fourfold.
However, within Type IIB, there is no such problem. The direction of
the inequality depends in an essential way on the dimension and, in
fact, $\chi(D,\Osheaf_D) > 0$ for an ample divisor in a Calabi-Yau
\emph{threefold}. In particular, condition eq.~\eqref{necessary} can
be already satisfied with smooth ample divisors.

We will thus consider a simple class of compactifications where the
superpotential is given by eq.~\eqref{eq:super} above and the K\"ahler
potential\footnote{Here we will ignore the complex structure and
  dilaton contributions ${\cal K}$ to the K\"ahler
  potential. Including such terms simply rescales the parameters of
  the superpotential by an overall multiplicative factor.} is given by
\begin{equation}
  \label{eq:kahler}
  K\big( T_i, \Tbar_i \big) 
  =
  -2 \ln V_X
  ,
\end{equation}
where the six-dimensional volume $V_X(\tau_i)$ is a homogeneous
function of degree $\tfrac{3}{2}$. Note that, while parameter $A$ in
eq.~\eqref{eq:super} is independent of the K\"ahler moduli, it does
depend on the complex structure moduli and the D3-brane positions if
D3-branes are present. Since the complex structure moduli are frozen
by the fluxes near the Kaluza-Klein (KK) scale, their dynamics is
essentially decoupled and one can treat $A$ as a constant parameter,
which can be discretely adjusted by scanning over the fluxes, just
like $W_0$. To further simplify the analysis, we will consider a setup
with no D3-branes. In addition, in the general case additional
non-perturbative terms are present in the superpotential, such as from
multiply wrapped D3-instantons. Thus, strictly speaking the single
term in eq.~\eqref{eq:super} should be regarded as the dominant
contribution in a series of contributions that determines the moduli
vevs, while the remaining terms are subleading and do not affect the moduli vevs. In
\autoref{consistency}, we demonstrate that such a truncation can be
made parametrically self-consistent, implying that the results
obtained are quite robust.

As will be shown in \autoref{minimize}, with the superpotential in
eq.~\eqref{eq:super} and the K\"ahler potential in
eq.~\eqref{eq:kahler}, it turns out that the supergravity scalar
potential has a supersymmetric Anti-de Sitter (AdS) extremum, the same
as in the KKLT proposal\footnote{There may also exist multiple Anti-de
  Sitter vacua with spontaneously broken supersymmetry similar to
  those found in~\cite{Acharya:2007rc}.  However, we restrict our
  discussion to the supersymmetric extremum because it is directly
  related to the de Sitter minimum obtained after the uplifting.}.  In
order to obtain de Sitter vacua, therefore, additional positive
contributions to the scalar potential are required. This can be
achieved either by introducing matter fields in the superpotential and
K\"ahler potential and including their dynamics via $F$ and
$D$-terms~\cite{Lebedev:2006qq, Achucarro:2006zf}, or by including
explicit supersymmetry violating terms in the potential, such as due
to adding a small number of anti D3-branes, which are trapped at the
bottom of the warped throat~\cite{Klebanov:2000hb} generated in these
flux compactifications.  The latter route was taken by
KKLT~\cite{Kachru:2003aw}, and for simplicity we will follow the same
approach. The potential is then given by
\begin{equation}
  \label{eq:scalarpot}
  V=
  e^K
  \left(
    G^{i\bar j} \, D_i W \, \overline{D_j W} 
    -3|W|^2
  \right)
  +
  \frac D{V_X^2}
  .
\end{equation} 
where $D$ is a constant. In a companion paper~\cite{PartTwo}, we study
alternative ways of obtaining de Sitter vacua.

\subsection{Coordinates on the K\"ahler Moduli Space}\label{coords}

In this section we discuss some properties of the K\"ahler moduli space
metric in Type IIB Calabi-Yau orientifolds. These properties are
completely general and are true for all Calabi-Yau threefolds. Therefore, 
they allow us to perform explicit computations without specifying a particular
manifold.

In mathematics, the K\"ahler moduli space is usually parametrized by
volumes $t_i$ of curves (two-cycles). Here,
$i\in\{1,\,\dots,\,\hplus{11}\}$ runs over a basis of
orientifold-invariant curves. Using these coordinates, the volume is a
cubic polynomial
\begin{equation}
  \label{eq:VXttt}
  V_X(t_i) = \sum d_{ijk} t_i t_j t_k
  ,
\end{equation}
where $d_{ijk}$ is the cup product\footnote{Using the Poincar\'e
  isomorphism $H^2(X,\Z)=H_4(X,\Z)$, the $d_{ijk}$ are the triple
  intersection number of the corresponding divisors.} of the basis
$(1,1)$-forms. However, physicists prefer different coordinates. They
parametrize the complexified K\"ahler moduli space $\MKahler$ of the
orientifold $X$ with holomorphic coordinates $T_i$ given by
\begin{equation}
  \label{eq:Tidef}
  T_i
  =
  \tau_i+i\chi_i
  ,\quad
  1\leq i \leq \hplus{11}(X)
  ,
\end{equation}
where $\tau_i$ are the volumes of four-cycles inside the manifold $X$
and $\chi_i$ are the axions corresponding to the periods of the
Ramond-Ramond four-form. As volumes of four-cycles, the $\{\tau_i\}$
variables are quadratic polynomials in the $\{t_i\}$. However, despite
the fact that we use the index ``$i$'' in both cases, the basis of
four-cycles is independent of the basis of two-cycles. A particularly
nice relative basis choice is where the cycles are Poincar\'e dual,
which amounts to setting
\begin{equation}
  \label{eq:dVdt}
  \tau_i 
  =
  \frac{\partial V_X}{\partial t_i}
  .
\end{equation}
We will always use this basis choice in the following. In order to
transform the two coordinates into each other we need the Jacobian
matrix, which is \emph{symmetric} because
\begin{equation}
  \label{eq:d2Vdt2}
  \frac{\partial \tau_i}{\partial t_j} 
  = 
  \frac{\partial^2 V_X}{\partial t_i \partial t_j}
  =
  \frac{\partial \tau_j}{\partial t_i}
  .
\end{equation}
Therefore,
\begin{equation}
  \label{eq:dVdtau}
  \frac{\partial V_X}{\partial \tau_i}
  =
  \sum_{j=1}^{\hplus{11}}
  \frac{\partial V_X}{\partial t_j}
  \frac{\partial t_j}{\partial \tau_i}
  =
  \sum_{j=1}^{\hplus{11}}
  \tau_j 
  \frac{\partial t_j}{\partial \tau_i}
  =
  \sum_{j=1}^{\hplus{11}}
  \tau_j 
  \frac{\partial t_i}{\partial \tau_j}
  =
  \frac{1}{2} t_i
  ,
\end{equation}
where we used Euler's homogeneous function theorem for $t_i$ being a
homogeneous function of the $\tau_j$ of degree $\tfrac{1}{2}$. The
homogeneity of $V_X$ and $\tau_i$ is the key property that will be
used in what follows.

The classical moduli space metric (not including possible quantum
corrections) is determined by the K\"ahler potential
\begin{equation}
  K = -2 \ln V_X 
  .
\end{equation}
Note that the dimensionless volume $V_X = \Vol(X)/\ell_s^6$ is a
homogeneous function of degree $\frac{3}{2}$ in the $\tau_i$, where
$\ell_s$ is the string scale. Let us define the following derivatives
with respect to the four-cycles\footnote{For reference we note that,
  using eq.~\eqref{eq:Tidef}, the derivatives are related by
  \begin{equation}
    \frac{\partial}{\partial \tau_i} 
    =
    \frac{\partial}{\partial T_i} +
    \frac{\partial}{\partial \Tbar_i} 
    ,\quad
    \frac{\partial}{\partial \chi_i} 
    =
    i \frac{\partial}{\partial T_i} -
    i \frac{\partial}{\partial \Tbar_i} 
    ,\quad
    \frac{\partial}{\partial T_i} 
    =
    \frac{1}{2} \frac{\partial}{\partial \tau_i} -
    \frac{i}{2} \frac{\partial}{\partial \chi_i} 
    ,\quad
    \frac{\partial}{\partial \Tbar_i} 
    =
    \frac{1}{2} \frac{\partial}{\partial \tau_i} +
    \frac{i}{2} \frac{\partial}{\partial \chi_i} 
    .
  \end{equation}
}
\begin{equation}
  \label{eq:Ki}
  K_i 
  =
  \frac{\partial K}{\partial \tau_i}
  =
  2 \frac{\partial K}{\partial T_i}
  =
  - \frac{2}{V_X} \frac{\partial V_X}{\partial \tau_i}
  =
  - \frac{t_i}{V_X}
  ,\quad
  K_{ij} 
  =
  \frac{\partial^2 K}{\partial \tau_i\partial \tau_j}
  =
  4
  \frac{\partial^2 K}{\partial T_i\partial{\Tbar}_{\bar j}}
  =
  4
  G_{i\bar j}
  ,
\end{equation}
where $G_{i\bar j}$ is the K\"ahler metric controlling the kinetic
terms in the Lagrangian. Since $V_X$ is a homogeneous function of
degree $\tfrac{3}{2}$, the first derivatives satisfy Euler's
homogeneous function theorem
\begin{equation}
  \label{eq:VXeuler}
  3 V_X = 
  \sum_{i=1}^{\hplus{11}} 
  \frac{\partial V_X}{\partial t_i}
  t_i 
  =
  \sum_{i=1}^{\hplus{11}} \tau_i  t_i 
  \qquad \Leftrightarrow \qquad
  \sum_{i=1}^{\hplus{11}} \tau_i  K_i = -3
  .
\end{equation}
This suggests to introduce new variables $a_i$ defined by
\begin{equation}
  \label{eq:aidef}
  a_i 
  = 
  - \frac{1}{2} \tau_i K_i
  = \frac{\tau_i \, t_i}{2 V_X} 
  ,
\end{equation}
with no sum over $i$ implied on the right hand side. Clearly, the
$a_i$ are of homogeneous degree $0$ in the $\tau_i$ (and the $t_i$),
decoupling the overall volume. In particular, we see from
eq.~\eqref{eq:VXeuler} that the $a_i$ satisfy
\begin{equation}
  \label{eq:aisum}
  \sum_{i=1}^{\hplus{11}}{a}_i
  =\frac{3}{2}
  .
\end{equation}

To summarize, there are three coordinate systems on the K\"ahler
moduli space:
\begin{enumerate}
\item The volumes $t_i$ of a fixed basis of curves
  (two-cycles). These have the advantage that the geometrically
  allowed values form a cone, the K\"ahler cone.
\item The volumes $\tau_i$ of a fixed basis of divisors or, more
  generally, four-cycles. These are the standard choice of fields in
  the supergravity action. 

  Thanks to the Hodge-Riemann bilinear relations $\det
  \big(\tfrac{\partial \tau}{\partial t}\big) >0$ for all $\vec{t}\in
  \Kc(X)$. Therefore, the basis change $\{t_i\} \leftrightarrow
  \{\tau_i\}$ is nowhere singular inside the K\"ahler cone, and both
  sets of fields are good coordinates on the moduli space.
\item The ``angular'' variables $a_i$. By construction, they do not
  scale with the overall size of the manifold, see
  eq.~\eqref{eq:aisum}. To parametrize the K\"ahler moduli space we
  need to pick a subset of $\hplus{11}-1$ variables $a_i$ plus one
  volume, which we take to be $V_X$.
\end{enumerate}

\subsection{Minimizing the Potential}\label{minimize}

Let us start with $D_iW=0$, the condition for unbroken
supersymmetry. Expanding the K\"ahler covariant derivative, one
obtains
\begin{equation}
  \begin{split}
    0
    =
    D_i W 
    =
    \frac{\partial W}{\partial T_i} + 
    W \frac{\partial K}{\partial T_i}
    =&\;
    -
    \frac{2\pi}N n_iAe^{-\frac{2\pi}{N}\vec n\cdot \vec T }
    + 
    \tfrac{1}{2} K_i
    \left(
      W_0 + A e^{-\frac{2\pi}{N} \vec{n}\cdot\vec{T}}
    \right)
    \\
    =&\;
    A e^{-\frac{2\pi}{N} \vec{n}\cdot\vec{T}}
    \bigg(
      \underbrace{
        - \frac{2\pi}{N} n_i + \frac{K_i}{2} 
      }_{\in \R_{<}}
      + 
      \frac{W_0\; e^{\frac{2\pi i}{N} \vec{n}\cdot\vec{\chi}} }{A} 
      \underbrace{
        \frac{K_i}{2} 
        e^{\frac{2\pi}{N} \vec{n}\cdot\vec{\tau}}
      }_{\in \R_{<}}
    \bigg)
    .
  \end{split}
\end{equation}
These are $\hplus{11}$ complex equations for the $2\hplus{11}$ real
variables $\{\tau_i,~\chi_i\}$. Note that $K_i$ is real and negative,
see eq.~\eqref{eq:Ki}. Therefore, in order to cancel, the third
summand must be real and positive, that is,
\begin{equation}
  \frac{1}{N}\; \vec{n}\cdot\vec{\chi}^\text{susy} + 
  \frac{1}{2\pi} \arg\left(\frac{W_0}{A}\right)
  \in 2 \Z + 1
  .
\end{equation}
In particular, we see that only the linear combination
$\vec{n}\cdot\vec{\chi}$ gets fixed to a specific value; If
$\tfrac{W_0}{A}$ has a complex phase then the axions dynamically
adjust. For simplicity, we will take $W_0$, $A$ to be real and of
opposite sign in the following, say, $W_0<0$ and $A>0$. Then one axion
will be stabilized at
\begin{equation}
  \vec{n}\cdot\vec{\chi}^\text{susy}=0 
  \mod
  \frac{N}{2\pi}
  .
\end{equation}
In fact, this feature is highly desirable if one tries to use the
axions in order to solve the strong CP problem in the context of
string theory. Having taken care of the axions, we are left with
$\hplus{11}$ real conditions for the variables $\{\tau_i\}$ or,
equivalently, $\{t_i\}$. Using eq.~\eqref{eq:Ki}, we obtain
\begin{equation}
  \label{eq:tisusy}
  \tsusy{i}
  =
  n_i
  \frac{
    \frac{4 \pi}{N} V_X
  }{
    \left( -\frac{W_0}{A}\right)
    e^{\frac{2\pi}{N} \vec{n}\cdot\vec{\tau}}
    -
    1
  }
  .
\end{equation}
We note that the supersymmetric
minimum is achieved at $\tsusy{i} \sim n_i$ with a positive
constant of proportionality. Therefore, the moduli are stabilized
within the K\"ahler cone, see eq.~\eqref{eq:Kc}, if and only if all
$n_i>0$. This is why the (single) divisor $D$ contributing to the
superpotential must be ample in order to stabilize the K\"ahler moduli
with all volumes positive.

So far we used the F-terms, but in reality we want to minimize the
scalar potential, eq.~\eqref{eq:scalarpot}, with the added
supersymmetry-violating term
\begin{equation}
  V
  =
  e^K
  \left(
    G^{i\bar j} \, 
    \frac{\partial W}{\partial T_i}\, 
    \overline{\frac{\partial W}{\partial T_j}} 
    -3|W|^2
  \right)
  +
  \frac{D}{V_X^2}
  .
\end{equation}
This is where the third parametrization of the K\"ahler moduli in
terms of, say, real coordinates $\big(a_1$, $a_2$, $\dots$,
$a_{\hplus{11}(X)-1}$, $V_X\big)$ plus their axionic partners becomes
essential. Note that the supersymmetry-breaking term $D/V_X^2$ depends
on $V_X$ only. Hence, the position of the minimum of the potential is
unchanged as far as the remaining K\"ahler moduli $a_i$ as well as all
axions are concerned. Therefore,
\begin{equation}
  \label{eq:aimin}
  \begin{split}
    \amin{i} =&\; \asusy{i}
    = 
    \frac{ 
      \tau_i\big( \vec{t}^\text{susy} \big) ~ \tsusy{i}
    }{ 
      2 V_X\big( \vec{t}^\text{susy} \big) 
    }
    = 
    \frac{ 
      n_i ~ \tau_i\big( \vec{n} \big)
    }{ 
      2 V_X\big( \vec{n} \big) 
    }
    = 
    n_i \frac{\partial}{\partial n_i} 
    \ln \sqrt{ \sum d_{jk\ell} n_j n_k n_\ell }
    ,
    \\
    \vec{n}\cdot \vec{\chi}^\text{min} 
    =&\; 
    \vec{n}\cdot \vec{\chi}^\text{susy} = 0
    ,
  \end{split}
\end{equation}
see eqns.~\eqref{eq:VXttt}, \eqref{eq:aidef}
and~\eqref{eq:tisusy}. Only the ``radial direction'' of the K\"ahler
moduli, parametrized by the volume $V_X$, is changed by the addition
of the supersymmetry-breaking term. Moreover, having fixed the $a_i$,
the volume $\tau_D=\vec{n}\cdot \vec{\tau}$ of the divisor $D$ is
proportional to $V_X^{2/3}$ with a constant of proportionality
\begin{equation}
  \label{eq:tauD_VX_prop}
  \frac{\tau_D}{\big(V_X\big)^{\frac{2}{3}}}
  =
  \frac{
    \sum n_i \, \tau_i\big( \vec{t}^\text{susy} \big) 
  }{
    V_X\big( \vec{t}^\text{susy} \big)^{\frac{2}{3}}
  }
  =
  \frac{
    \sum n_i 
    \frac{\partial}{\partial n_i}
    \big( \sum d_{jk\ell} n_j n_k n_\ell \big)
  }{
    \big( \sum d_{jk\ell} n_j n_k n_\ell \big)^{\frac{2}{3}}
  }
  =
  3
  \bigg( \sum_{jk\ell} d_{jk\ell} n_j n_k n_\ell  \bigg)^{\frac{1}{3}}
  .
\end{equation}

Therefore, minimizing the scalar potential for the $\hplus{11}$ fields
$\{T_i\}$ boils down to minimizing it with respect to the single
complexified K\"ahler modulus
\begin{equation}
  T_D 
  =
  \tau_D + i \chi_D
  =
  (\vec{n}\cdot\vec{\tau}) + i \; (\vec{n}\cdot\vec{\chi})
  ,
\end{equation}
where we moreover already know that $\chisusy{D}=\chimin{D}=0\mod
\tfrac{N}{2\pi}$, though we will not eliminate it quite yet. The
scalar potential simplifies to
\begin{equation}
  \label{eq:PotD}
  \begin{split}
    V 
    =&\;
    \frac{1}{V_X^2(\tau_D)}
    \left[
      \Big(\tfrac{1}{4}\frac{\partial^2 K}{\partial \tau_D^2}\Big)^{-1}
      | D_{T_D}W |^2
      - 3 |W|^2 
    \right]
    +
    \frac{D}{V_X^2(\tau_D)}
    \\
    =&\;
    \frac{1}{V_X^2(\tau_D)}
    \left[
      \frac{8\pi}{N} \tau_D 
      A e^{-\frac{2\pi}{N} \tau_D}
      \left(
        \big( \tfrac{2\pi}{3 N} \tau_D + 1 \big)
        A e^{-\frac{2\pi}{N} \tau_D}
        + W_0 \cos\Big(\frac{2\pi \chi_D}{N}\Big)
      \right)
      +
      D
    \right]
    ,
  \end{split}
\end{equation}
where we used eqns.~\eqref{eq:Ki} and
\begin{equation}
  \label{eq:TDsusy}
  D_{T_D}W
  =
  \frac{1}{2} 
  \Big(
    D_{\tau_D}W - i D_{\chi_D}W
  \Big)
  =
  - \frac{1}{2 \tau_D} 
  \left(
    \big(\tfrac{4\pi}{N} \tau_D +3\big) 
    A e^{-\frac{2\pi}{N} (\tau_D + i \chi_D)}
    + 3 W_0
  \right)
  .
\end{equation}
Minimizing with respect to $\chi_D$, we immediately recognize that
\begin{equation}
  \frac{\partial V}{\partial \chi_D}
  \stackrel{!}{=}
  0
  \quad\Leftrightarrow\quad
  \sin\Big(\frac{2\pi \chimin{D}}{N}\Big) = 0
  \quad\Leftrightarrow\quad
  \chimin{D}=\chisusy{D}=0
  \mod
  \tfrac{N}{2\pi}
\end{equation}
and the addition of the supersymmetry-breaking term does, indeed, not
change the axions. Hence we set $\chi_D=0$ and proceed to minimize
with respect to $\tau_D$,
\begin{equation}
  \label{eq:VDmin}
  \frac{\partial V}{\partial \tau_D} 
  = 
  \frac{32 \pi}{3 N}  
  \frac{\tau_D}{V_X^2}
  \Big(\frac{\pi}{N} \tau_D + 1\Big)
  \Big(D_{T_D} W\big|_{T_D=\tau_D} \Big)
  A e^{-\frac{2\pi}{N} \tau_D}
  -
  \frac{3 D}{V_X^2 \tau_D}
  \stackrel{!}{=}
  0
  .
\end{equation}
As expected, if we set $D=0$ then supersymmetry is restored and the
minimum of the scalar potential is given by the F-term
$D_{T_D}W=0$. However, if we turn on $D\not=0$, the stabilized volume
$V_D^\text{min}$ is implicitly determined by the transcendental
eq.~\eqref{eq:VDmin}, but cannot be solved for analytically. In the
remainder of this section, we will compare it to the supersymmetric
solution
\begin{equation}
  \label{eq:VDsusy}
  0 
  \stackrel{!}{=}
  \Big( D_{T_D}W \Big) \Big|_{T_D = \taususy{D}}
  \qquad \Leftrightarrow \qquad
  \taususy{D} = 
  \frac{N}{2\pi}
  \left[
    - \Omega_{-1}\left(\frac{3 W_0}{2 A} e^{-\frac{3}{2}} \right) 
    - \frac{3}{2}
  \right],
\end{equation}
where we remind the reader that $W_0/A<0$ in our
notation. Here\footnote{We will use $\Omega$ instead of $W$ for the
  Lambert W-function to differentiate it from the superpotential.}
$\Omega_{-1}$ is the non-principal branch of the Lambert
W-function. It is real and negative on the domain
\begin{equation}
  - e^{-1}
  < 
  \frac{3 W_0}{2 A} e^{-\frac{3}{2}} 
  < 
  0
  ,
\end{equation}
with an essential singularity at $0$. Using the expansion
\begin{equation}
  \Omega_{-1}(x) = \ln(-x) - \ln\big(-\ln(-x)\big) + \dots
\end{equation}
around $x=0^-$, we obtain
\begin{equation}
  \taususy{D}
  =
  \frac{N}{2\pi}
  \bigg(
    \ln \left|\tfrac{2 A}{3 W_0} \right| 
    + \ln\Big(\tfrac{3}{2}+\ln\left|\tfrac{2 A}{3 W_0}\right| \Big)
    + \dots
  \bigg)
  .
\end{equation}
For the above solution to be valid, that is, to be in the regime in
which the supergravity approximation holds, requires
\begin{equation}
  \label{sugraapprox}
  0 
  < 
  - \frac{3 W_0}{2 A}
  \ll
  1
  \quad\Rightarrow\quad
  \left|\frac{2 A}{3 W_0}\right|
  \gg 
  1
  .
\end{equation} 
Taking $A$ to be ${\cal O}(1)$, this automatically implies that the
gravitino mass\footnote{This is true if the volume $V_X$ is consistent
  with standard gauge unification, i.e. it is not too large.}
$m_{3/2}= e^{K/2}\,W \approx W_0$ is much smaller than the Planck
scale\footnote{Here we set $m_\text{Pl}=1$.}, and hence could give
rise to low energy supersymmetry at around the TeV scale.

To estimate the effect of a small supersymmetry breaking term, we
use eq.~\eqref{eq:VDsusy} to Taylor expand
\begin{equation}
  \left.
    \frac{\partial V}{\partial \tau_D} 
  \right|_{\tau_D=\taususy{D} + \epsilon}
  =
  \frac{-3 D}{\taususy{D} \; V_X^2(\taususy{D})}
  +
  12
  \frac{
    \tfrac{8\pi^2}{N^2}
    \big( \frac{\pi}{N} \taususy{D} +  1 \big)
    \big( \frac{4 \pi}{N} \taususy{D} +  1 \big)
    \big(\taususy{D} W_0 \big)^2
  }{
    \big(\tfrac{4\pi}{N} \taususy{D} + 3\big)^2
    \big(\taususy{D}\big)^2 \; 
    V_X^2(\taususy{D})
  }
  \epsilon
  + 
  O(\epsilon D, \epsilon^2)
  .
\end{equation}
Therefore, to first order, the minimum of the scalar potential is at
\begin{equation}\label{eq:tauDmin}
  \taumin{D}
  = 
  \taususy{D} +
  \frac{N^2}{32 \pi^2}
  \frac{
    \big( \frac{4\pi^2}{N^2} \taususy{D} + 3 \big)^2
  }{
    \big( \frac{\pi}{N} \taususy{D} +  1 \big)
    \big( \frac{4 \pi}{N} \taususy{D} +  1 \big)
    \taususy{D}
    W_0^2
  }
  D
  + 
  O(D^2)
  .
\end{equation}
To summarize, we have found that:
\begin{itemize}
\item The real part of the K\"ahler moduli are stabilized inside the
  K\"ahler cone, that is, with all volumes positive.
\item One axion is stabilized, the remaining $\hplus{11}-1$ remain
  massless.
\item The only effect of the supersymmetry-breaking addition to the
  scalar potential is to increase the overall volume at which the
  moduli are fixed. Neither the axions nor the angular part of the
  K\"ahler moduli is changed. 
\end{itemize}

Finally, before moving on to providing explicit examples of Calabi-Yau
orientifolds, which realize the above mechanism, it is worthwhile to
comment on the size of the explicit supersymmetry breaking
contribution due to anti D3-branes. As mentioned earlier, these
positive contributions to the potential can give rise to a de Sitter
vacuum under certain conditions, the same as in the KKLT
proposal. These contributions are exponentially small due to strong
warping at the bottom of the warped throat where these anti D3-branes
are dynamically trapped, leading to $0<D\ll 1$. Therefore, to obtain a
de Sitter vacuum this requires the tree-level superpotential
contribution $W_0$ to be exponentially small as well since otherwise
the uplifting would not be strong enough and the vacuum would remain
AdS. Thus, we see that the requirement of a small $W_0 \ll 1$ has
three separate origins:
\begin{itemize}
\item Moduli Stabilization in the supergravity regime.
\item A small gravitino mass relative to the Planck scale.
\item Realizing a de Sitter Vacuum.
\end{itemize}
As mentioned earlier, although one naturally expects $W_0={\cal
  O}(1)$, by choosing appropriate values of the integer fluxes it is
possible to tune $W_0 \ll 1$, see~\cite{Giryavets:2003vd, Giryavets:2004zr}. Then, as we
will explicitly see in \autoref{sec:numerical}, one can actually
arrange the numerical values to yield a de Sitter vacuum with a small
cosmological constant.

\section{Explicit Examples}
\label{sec:example}

In this section, we provide some explicit examples of Calabi-Yau
orientifolds. To start with, we will consider a simple case with just
one K\"ahler modulus and construct a divisor which is both rigid and
ample, thereby contributing to the superpotential. We then consider
the relevant case with three K\"ahler moduli in which we realize the
mechanism outlined in \autoref{sec:single}.

\subsection{Simple Case - A One-Parameter Model}
\label{simplecase}

The simplest way for D3-instantons to stabilize the K\"ahler moduli in
the interior of the K\"ahler cone is to have a single instanton
wrapped on a rigid ample divisor. Recall from \autoref{super} that
this automatically guarantees that $\chi_+-\chi_-=1$ and, therefore,
the divisor contributes to the superpotential.

The starting point for our example is the $\CP^7[2222]$ complete
intersection Calabi-Yau manifold, that is, the complete intersection
of $4$ quadrics in $\CP^7$. There is a free $\Z_2\times\Z_2\times
\Z_2$-action on $\Xt$~\cite{MR2373582, hua-2007, SHN, Braun:2009qy,
  Classification}. We will always divide out this group action and
define
\begin{equation}
  X = \Xt \big/\big( \Z_2\times \Z_2 \times \Z_2 \big)
  .
\end{equation}
The Hodge numbers of the CICY threefold $\Xt$ are well-known. Since
the $(\Z_2)^3$ action necessarily preserves the K\"ahler class, we
obtain
\begin{equation}
  h^{pq}\big(\Xt\big) 
  =
  \vcenter{\xymatrix@!0@=5mm@ur{
      1 &  0 &  0 & 1 \\
      0 & 65 &  1 & 0 \\
      0 &  1 & 65 & 0 \\
      1 &  0 &  0 & 1 
    }}
  ,\qquad
  h^{pq}\big(X\big) 
  =
  \vcenter{\xymatrix@!0@=5mm@ur{
      1 &  0 &  0 & 1 \\
      0 &  9 &  1 & 0 \\
      0 &  1 &  9 & 0 \\
      1 &  0 &  0 & 1 
    }}
  .
\end{equation}
The group action on the homogeneous coordinates $z_0$, $\dots$, $z_7$
of $\CP^7$ is the \emph{regular representation}. Explicitly, it is
generated by
\begin{equation}
  \begin{split}
    g_1 
    =&\;
    \diag(+1,+1,+1,+1,-1,-1,-1,-1)
    ,\\
    g_2
    =&\;
    \diag(+1,+1,-1,-1,+1,+1,-1,-1)
    ,\\
    g_3
    =&\;
    \diag(+1,-1,+1,-1,+1,-1,+1,-1)
    .
  \end{split}
\end{equation}
A basis for the $(\Z_2)^3$-invariant polynomials is spanned by $p_i =
z_i^2$, $0\leq i \leq 7$. One can check~\cite{GPS05} that the zero set
of four generic linear combinations of invariant degree-$2$
polynomials is a \emph{smooth} Calabi-Yau threefold $\Xt$, and that
the $(\Z_2)^3$ action on $\Xt$ is fixed-point free. Therefore, the
quotient $X$ is again a smooth Calabi-Yau threefold.

We now consider the line bundle $\Osheaf_\Xt(1)$, whose sections are 
\begin{equation}
  H^0\big( \Xt, \Osheaf_\Xt(1) \big) = 
  \Span_{\C} \{ z_0,~\dots,~ z_7 \} 
  ,
\end{equation}
and all higher-degree cohomology groups vanish. Decomposing this
representation into irreducible representations, we find a unique
invariant section
\begin{equation}
  \label{eq:Dsections}
  H^0\big(X, \Osheaf_X(1) \big) = 
  H^0\big( \Xt, \Osheaf_\Xt(1) \big)^{\Z_2^3} = 
  \C \cdot z_0.
\end{equation}
We denote the corresponding divisor by
\begin{equation}
  \Dt =
  \big\{ z_0 = 0 \big\} \subset \Xt 
  ,\quad
  D = \Dt \big/(\Z_2^3)
  \subset X
  .
\end{equation}
The Hodge numbers of the divisors are related to the line bundle
cohomology by
\begin{equation}
  \label{eq:hodgeD}
  h^{00}(D) = h^2\big( X, \Osheaf(D) \big) + 1
  ,\quad
  h^{01}(D) = h^1\big( X, \Osheaf(D) \big) 
  ,\quad
  h^{02}(D) = h^0\big( X, \Osheaf(D) \big) -1
  .
\end{equation}
They enjoy the following properties:
\begin{itemize}
\item Both $\Dt$, $D$ are ample because $\Osheaf(1)$ is an ample
  line bundle.
\item $D$ is rigid by eq.~\eqref{eq:Dsections}. 
\item Using~\cite{GPS05}, one can check that $D$ is a \emph{smooth}
  complex surface for generic complex structure moduli of $X$.
\end{itemize}
Therefore, $D$ satisfies the sufficient requirements to contribute to
the superpotential.

Let us further investigate the geometry of the divisor $D$. A standard
computation yields the Chern numbers. By abuse of notation, we denote
by $h$ the hyperplane classes in $\CP^7$ as well as its restriction to
$\Xt$, $\Dt$. Using this notation,
\begin{equation}
  \begin{aligned}
    c_1\big(\Xt\big) =&\; -K_\Xt = 0
    ,
    &
    c_2\big(\Xt\big) =&\; 4 h^2
    ,
    \qquad&
    c_3\big(\Xt\big) =&\; -8 h^3
    ,
    \\
    c_1\big(\Dt\big) =&\;  -K_\Dt = -h
    ,
    \qquad &
    c_2\big(\Dt\big) =&\; 5 h^2
    .
    \qquad &
  \end{aligned}
\end{equation}
It is now easy to compute the Chern numbers, and we obtain
\begin{equation}
  \begin{split}
    \int c_1\big(\Dt\big)^2 = 16
    \qquad&\Rightarrow~
    \int c_1(D)^2 = 2
    ,
    \\
    \chi(\Dt) = 
    \int c_2\big(\Dt\big) = 80
    \qquad&\Rightarrow~
    \int c_2(D) = 10
    = \chi(D)
    .
  \end{split}
\end{equation}
Using the Euler number, and the above properties of the divisors, it
is easy to determine their Hodge numbers,
\begin{equation}
  h^{pq}\big(\Dt\big) 
  =
  \vcenter{\xymatrix@!0@=5mm@ur{
      7 &  0 & 1 \\
      0 & 64 & 0 \\
      1 &  0 & 7 \\
    }}
  ,\qquad
  h^{pq}\big(D\big) 
  =
  \vcenter{\xymatrix@!0@=5mm@ur{
      0 &  0 & 1 \\
      0 &  8 & 0 \\
      1 &  0 & 0 \\
    }}
  .
\end{equation}
We observe that $D$ is a numerical Campedelli
surface~\cite{BeauvilleNonAbel, MR2390332} with
$\pi_1(D)=\Z_2\times\Z_2\times\Z_2$.

In fact, the appearance of this surface is not a coincidence. By
adjunction, an ample divisor in a Calabi-Yau threefold is a complex
surface with ample canonical\footnote{Not to be confused with del
  Pezzo surfaces, whose \emph{anti}-canonical bundle is ample.}
bundle. Such a surface is called \emph{of general type}, and can be
roughly classified by the Chern numbers $c_1^2 = \int c_1(D)^2$ and
$c_2 = \int c_2(D)$. If we now furthermore impose the arithmetic genus
$\chi(D,\Osheaf_D)=1$, then Hirzebruch-Riemann-Roch yields
\begin{equation}
  1 = \chi(D,\Osheaf_D) = \int \ch(\Osheaf_D) \Td(TD) =
  \int \frac{c_1(D)^2+c_2(D)}{12}
  .
\end{equation}
Since $c_1^2$, $c_2>0$ for surfaces of general type, this leaves us
with $11$ possibilities. The Bogomolov-Miyaoka-Yau inequality
$c_1^2\leq 3 c_2$ excludes two. 
\begin{table}
  \centering
  \begin{tabular}{|c@{\quad}c|@{\quad}c@{\quad}l@{\quad}l|}
    \hline 
    $c_1^2$& $c_2$& $|\pi_1(D)|<\infty$& 
    Example & Ambient Calabi-Yau \\
    \hline\hline
    $1$& $11$& Yes & Godeaux surface & Quintic$\big/ \Z_5$ \\
    $2$& $10$& Yes & Campedelli surfaces & 
    CICY $\#7884\big/(\Z_3\times\Z_3)$ \\
    $3$& $9$ & Yes & Burniat surfaces &
    CICY $\#7862\big/(\Z_2\times Q_8))$\\
    $4$& $8$ & Yes & Burniat surfaces & CICY
    $\#7861\big/(\Z_8\rtimes\Z_4)$ \\
    $5$& $7$ & Yes & Burniat surfaces & ? \\
    $6$& $6$ & No  & Burniat surfaces & N/A \\
    $7$& $5$ & No  & Inoue surface & N/A \\
    $8$& $4$ & No  & Beauville surface & N/A\\
    $9$& $3$ & No  & fake $\CP^2$ & N/A
    \\  \hline
  \end{tabular}
  \parbox{12cm}{
    \caption{Examples of surfaces of general type with
      $\chi(D,\Osheaf_D)=1$ and how they arise as rigid ample divisors
      in Calabi-Yau threefolds.}  
    \label{tab:generaltype}
  }
\end{table}
We list a number of examples in \autoref{tab:generaltype}. Note that
the surfaces arising as ample divisors in Calabi-Yau manifolds have
$|\pi_1(D)|=|\pi_1(X)|< \infty$, which further restricts their Chern
numbers to $1\leq c_1^2\leq 5$.

So far, we have not specified the orientifold involution. The obvious
choice is to pick one homogeneous coordinate $z_j$ and send
$z_j\mapsto -z_j$. This is a well-defined orientifold action on $X$
because $\Xt$ is invariant and the orientifold action commutes with
the $(\Z_2)^3$-action. The fixed-point set is a single O7-plane. We
distinguish the following two cases:
\begin{itemize}
\item $j=0$: $D$ sits on top of the $O7$.
\item $j\not=0$: $D$ intersects the $O7$-plane in a genus-$3$ curve.
\end{itemize}
In either case, a D3-instanton wrapped on the divisor $D$ satisfies
the necessary criteria to contribute to the superpotential.

\subsection{A Three-Parameter Model}
\label{mainexample}

\subsubsection{The Geometry}

We now consider a particular example of a Calabi-Yau threefold
with $h^{11}=h^{21}=19$.  It contains an ample divisor $D$ with $\vec n=\{1,\,1,\,1\}$ such that
$\chi_{+}-\chi_{-}=1$. Note that $D$ is not rigid, yet still satisfies
the necessary condition to contribute to the superpotential. We start with the CICY \#18,
\begin{equation}
  \begin{array}{r@{}}
    \CP^1 \\ \CP^2 \\ \CP^3
  \end{array}
  \!
  \left[
    \begin{array}{ccccc}
      0 & 1 & 1 \\
      0 & 0 & 3 \\
      3 & 1 & 0
    \end{array}
  \right]
  .
\end{equation}
Now, define the orientifold $\Z_2$-action to be generated by
\begin{equation}
  \label{eq:OmegaAct}
  \Omega 
  \Big(
  \big[
  x_0:x_1
  \big|
  y_0:y_1:y_2
  \big|
  z_0:z_1:z_2:z_3
  \big]
  \Big)
  =
  \big[
  x_0:-x_1
  \big|
  y_0:-y_1:-y_2
  \big|
  z_0:-z_1:-z_2:-z_3
  \big]
\end{equation}
and demand that the polynomials (of the required multi-degrees)
transform as
\begin{equation}
  p_{(0,0,3)}\circ \Omega
  =
  -p_{(0,0,3)}
  ,\quad
  p_{(1,0,1)}\circ \Omega
  =
  p_{(1,0,1)}
  ,\quad
  p_{(1,3,0)}\circ \Omega
  =
  p_{(1,3,0)}
  .
\end{equation}
Note that, although these polynomials are not invariant, their zero
set is. Using~\cite{GPS05}, one can check that generic such
polynomials cut out a smooth\footnote{We remark that
  eq.~\eqref{eq:OmegaAct} is the unique action that admits a smooth
  embedded threefold.} Calabi-Yau threefold $X=\big\{p_{(0,0,3)}=
p_{(1,0,1)}= p_{(1,3,0)}=0\big\}$. Moreover, a generic
$\Z_2$-invariant polynomial
\begin{equation}
  p_{(1,1,1)}\in H^0\big(X, \Osheaf(1,1,1)\big)^{\Z_2}
\end{equation}
defines a smooth divisor $D=\big\{p_{(1,1,1)}=0\big\} \subset X$,
which we will take to be our D3-instanton. Using the Koszul spectral
sequence, one can compute that
\begin{equation}
  \chi_+\big( D, \Osheaf(D) \big) = 11
  ,\quad
  \chi_-\big( D, \Osheaf(D) \big) = 10
  .
\end{equation}
The divisor $D$ is ample, and therefore is a surface of general type
with $\pi_1(D)=\pi_1(X)=1$. Its Chern numbers are $c_1^2=90$, $c_2=162$.
\begin{table}
  \centering
  \renewcommand{\arraystretch}{1.5}
  \begin{tabular}{|@{\quad}ccc@{\quad}|@{~}cccc@{~}|@{$\quad$}l@{$\quad$}l@{\quad}|}
    \hline
    $\CP^1$ & $\CP^2$ & $\CP^3$ & 
    \begin{sideways}$p_{(0,0,3)}\equiv 0~$\end{sideways} & 
    \begin{sideways}$p_{(1,0,1)}\equiv 0$\end{sideways} & 
    \begin{sideways}$p_{(1,3,0)}\equiv 0$\end{sideways} & 
    \begin{sideways}$p_{(1,1,1)}\equiv 0$\end{sideways} & 
    O-planes & O-planes $\cap D$ \\
    \hline\hline
    $[1:0]$ & $[0:*:*]$ &  $[0:*:*:*]$ & 
    N & Y & Y & N & O7 & $T^2$ \\
    $[0:1]$ & $[0:*:*]$ &  $[0:*:*:*]$ & 
    N & N & N & Y & Nine O3 & $9$ points \\
    $[1:0]$ & $[1:0:0]$ &  $[0:*:*:*]$ & 
    N & Y & N & Y & --- & --- \\
    $[0:1]$ & $[1:0:0]$ &  $[0:*:*:*]$ & 
    N & N & Y & N & Three O3 & --- \\    
    $[1:0]$ & $[1:*:*]$ &  $[1:0:0:0]$ & 
    Y & N & Y & Y & --- & --- \\
    $[0:1]$ & $[1:*:*]$ &  $[1:0:0:0]$ & 
    Y & Y & N & N & Three O3 & --- \\    
    $[1:0]$ & $[1:0:0]$ &  $[1:0:0:0]$ & 
    Y & N & N & N & --- & --- \\
    $[0:1]$ & $[1:0:0]$ &  $[1:0:0:0]$ & 
    Y & Y & Y & Y & O3 & 1 point \\    
    \hline
  \end{tabular}
  \parbox{12cm}{
    \caption{Orientifold planes on $X\subset
      \CP^1\times\CP^2\times\CP^3$ arising from different patches of the
      ambient space. Geometrically, the O7-plane spans $\R^{3,1}\times
      \CP^1\times T^2$.}
    \label{tab:oplanes}
  }
\end{table}
It intersects the O7-plane in a smooth elliptic curve as well as $10$
out of the $16$ O3-planes, see \autoref{tab:oplanes}. Therefore, the
Hodge numbers of $D$ are
\begin{equation}
  h^{pq}(D) 
  =
  \vcenter{\xymatrix@!0@=5mm@ur{
      21 &  0 & 1 \\
      0 & 118 & 0 \\
      1 &  0 & 21 \\
    }}
  ,\qquad
  \hplus{pq}(D) 
  =
  \vcenter{\xymatrix@!0@=5mm@ur{
      10 &  0 & 1 \\
      0 & 64 & 0 \\
      1 &  0 & 10 \\
    }}
  ,\qquad
  \hminus{pq}(D) 
  =
  \vcenter{\xymatrix@!0@=5mm@ur{
      10 &  0 & 0 \\
      0 & 56 & 0 \\
      0 &  0 & 10 \\
    }}
  .
\end{equation}
The Calabi-Yau threefold has Hodge numbers $h^{11}(X)=19=h^{21}(X)$,
but only a $3$-dimensional sublattice of $H_4(X,\Z)\simeq \Z^{19}$ is
spanned by the divisors $h_1=\{x_0=0\}$, $h_2=\{y_0=0\}$,
$h_3=\{z_0=0\}$. As long as we are only considering instantons wrapped
on divisors $D$ in this sublattice, we can consistently ignore the
remaining $19-3=16$ K\"ahler moduli. Therefore, we parametrize the
K\"ahler class as $\omega = t_1 h_1 + t_2 h_2 + t_3 h_3$. The volumes
of the relevant (sub-)manifolds are\footnote{Note that, as an abstract
  cubic polynomial in three variables, $V_X$ has discriminant
  $\Delta=0$. This proves that no coordinate change can bring it into
  the swiss cheese form, which would have $\Delta\not=0$.}
\begin{equation}
  \label{eq:tis}
  \begin{split}
    V_X =&\; \tfrac{3}{2} t_2 t_3 (6 t_1+t_2+3 t_3)
    ,\\
    \tau_1 = \frac{\partial V_X}{\partial t_1} =
    \Vol(h_1) =&\; 9 t_2 t_3
    ,\\
    \tau_2 = \frac{\partial V_X}{\partial t_2} =
    \Vol(h_2) =&\; \tfrac{3}{2} t_3 (6 t_1+2 t_2+3 t_3)
    ,\\
    \tau_3 = \frac{\partial V_X}{\partial t_3} =
    \Vol(h_3) =&\; \tfrac{3}{2} t_2 (6 t_1+t_2+6 t_3)
    ,\\
    \tau_D = \vec{n}\cdot \vec{\tau}
    =&\;
    \tfrac{9}{2} t_3^2+\tfrac{3}{2} t_2^2+
    9 t_1 t_2+ 9 t_1 t_3+ 21 t_2 t_3
    .
  \end{split}
\end{equation}
In these coordinates, the K\"ahler cone is precisely the first octant
$t_1$, $t_2$, $t_3>0$.

\subsubsection{Stabilization of Moduli}
\label{modulivevs}

Armed with an expression for the volume in terms of the two-cycles
$t_i$, and that for the $t_i$ in terms of the $\tau_i$ in
eq.~\eqref{eq:tis}, we can express the volume $V_X$ in terms $\tau_1$,
$\tau_2$ and $\tau_3$. As we remarked earlier, the coordinate change
is one-to-one inside the K\"ahler cone. However, since the $\tau_i$
are quadratic in the $t_i$, some non-physical values of the $t_i$ are
also mapped to allowed values of the $\tau_i$. Hence, when we invert
the quadrics, care must be taken to choose the correct branch for each
root. The unique solution for the two-cycles in terms of the
four-cycles is
\begin{equation}
  \label{eq:t_tau}
  t_1=
  \frac{
    4(3\tau_2-\tau_1)\tau_3-3\tau_1(4\tau_2-\tau_1)
  }{
    6\sqrt{3}\sqrt{\tau_1(6\tau_2-\tau_1)(2\tau_3-\tau_1)}
  }
  , \quad
  t_2=
  \frac{
    \sqrt{\tau_1(2\tau_3-\tau_1)}
  }{
    \sqrt 3\sqrt{(6\tau_2-\tau_1)}
  }
  , \quad
  t_3=
  \frac{
    \sqrt{\tau_1(6\tau_2-\tau_1)}
  }{
    3\sqrt 3\sqrt{(2\tau_3-\tau_1)}
  }
  .
\end{equation}
Then, the K\"ahler potential equals
\begin{equation}
  \label{eq:kpoten}
  K 
  = 
  -2\ln\Big(
    \tfrac{3}{2} t_2 t_3 (6 t_1+t_2+3 t_3)
  \Big)
  =
  -2\ln\left(
    \frac 1{6\sqrt 3}
    \sqrt{\tau_1(6\tau_2-\tau_1)(2\tau_3-\tau_1)}
  \right)
  .
\end{equation}
There exists an ample divisor satisfying the criterion
$\chi_+-\chi_-=1$, whose volume is given by
\begin{equation}
  \tau_D=\tau_1+\tau_2+\tau_3
  ,
\end{equation}
that is, $n_1=n_2=n_3=1$. Therefore, we can now directly apply the
methods developed in the previous subsection. The ``angular'' part of
the moduli vevs is
\begin{equation}
  \amin{1} = \asusy{1}
  =\frac{6}{20}
  ,\qquad
  \amin{2} = \asusy{2}
  =\frac{11}{20}
  ,\qquad
  \amin{3} = \asusy{3}
  =\frac{13}{20}
  ,
\end{equation}
using eq.~\eqref{eq:aimin}.     

The two alternative choices for coordinates on
the moduli space, $\{t_i\}$ and $\{\tau_i\}$, are
stabilized in the directions $t_i\sim n_i$ and $\tau_i \sim
\tfrac{a_i}{n_i}$ respectively, see eqns.~\eqref{eq:tisusy}
and~\eqref{eq:aidef}. The constant of proportionality can be
parametrized by one volume, which we take to be $\tau_D$. Using
eq.~\eqref{eq:VXttt} and $\tau_D=\vec{n}\cdot \vec{\tau}$, we can fix
the constant of proportionality in general to be
\begin{equation}
  \label{eq:tauitauDprop}
  t_i =
  \left(
    \frac{V_X}{\sum d_{jk\ell} n_j n_k n_\ell}
  \right)^{\frac{1}{3}}
  n_i
  ,\qquad
  \tau_i = 
  \frac{2}{3} \frac{a_i}{n_i} \tau_D
  .
\end{equation}
For the case at hand, we obtain 
\begin{equation}
  \label{eq:tauandt}
  \tau_1=\frac{6}{30} \tau_D
  ,\quad
  \tau_2=\frac{11}{30} \tau_D
  ,\quad
  \tau_3=\frac{13}{30} \tau_D
  ,\quad
  t_1=t_2=t_3=
  \frac{\sqrt{\tau_D}}{3\sqrt{5}} 
  ,\quad
  V_X=\frac{\tau_D^{3/2}}{9\sqrt{5}}
  ,
\end{equation}
using eq.~\eqref{eq:t_tau} and~\eqref{eq:tauD_VX_prop}. 

Note that in this explicit example all the four-\ and two-cycle
volumes as well as the volume of the Calabi-Yau are functions of a
single parameter - the volume $\tau_D$ of the ample divisor, which is determined by $W_0$, $A$, $N$ and $D$, 
and is in principle computable from the choice of fluxes and anti-D3
branes. This implies that in a compactification with a realistic  
visible sector, the volume of the visible sector four-cycle, which measures the gauge coupling 
at the compactification (KK) scale is also fixed in terms of $\tau_D$. Thus, choosing 
phenomenologically well motivated values of 
$\tau_{\rm  visible}$ at $M_{KK}\sim M_{GUT}$, such as $\tau_{\rm
  visible}\approx \alpha_{GUT}^{-1}\approx 25$, completely determines the values of all
K\"ahler moduli, in particular the volume of the Calabi-Yau
manifold! Furthermore, it provides a bottom-up constraint on the microscopic parameters $W_0,\,A,\,N$, and $D$. 
Of course, the complex structure moduli and the dilaton, whose vevs are controlled by the fluxes remain far less
constrained\footnote{In contrast, consider the fluxless $G_2$
  compactifications of $M$-theory~\cite{Acharya:2008hi}. Here, one actually constrains
  \emph{all} moduli (three-\ and four-cycles) in terms of a single
  parameter --- the volume $V_{\cal Q}$ of an associative three-cycle
  ${\cal Q}$ that intersects all co-associative four-cycles positively}.

\subsubsection{Supersymmetry Breaking and a Particular Choice of
  Parameters}
\label{sec:numerical}

\begin{figure}
  \centering
  \input{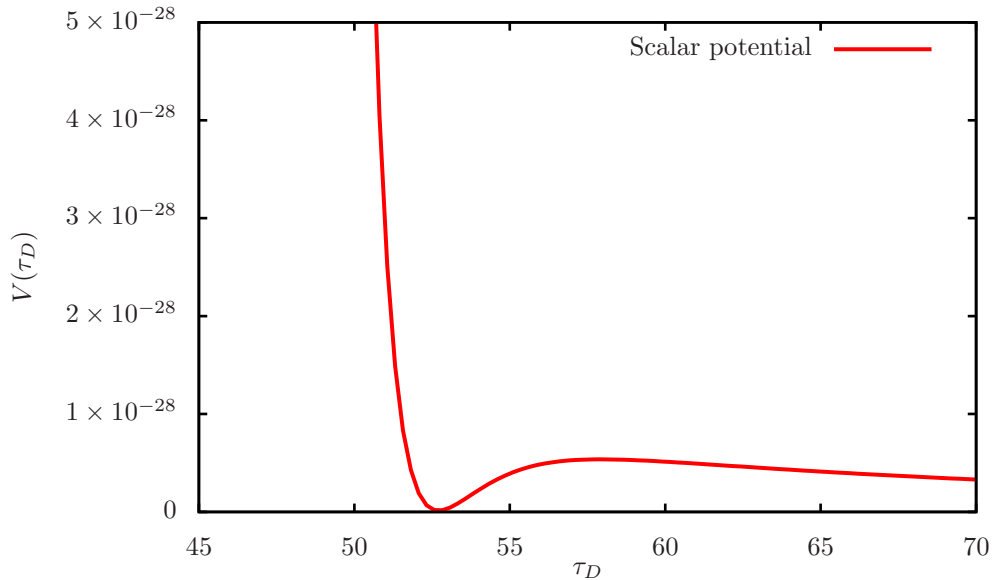}
  \caption{The scalar potential in the $\tau_D$-direction for the
    parameter choice in eq.~\eqref{eq:sample}.}
  \label{fig:scalarpot}
\end{figure}

As a concrete numerical example, consider the following choice of input parameters for 
the three parameter model considered in \autoref{mainexample}
\begin{equation}
  \label{eq:sample}
  A=1
  ,\quad
  W_0=-10^{-13}
  ,\quad
  N=10
  ,\quad
  D=2.8\times 10^{-26}
  .
\end{equation}
Here we have assumed the origin of the non-perturbative superpotential
to be a gaugino condensate with a dual Coxeter number $N=10$ as this
helps in stabilizing the moduli in the supergravity regime. Also, as
mentioned earlier, both $W_0$ and $D$ are required to be extremely
small for the solution to be self-consistent \emph{and} be
phenomenologically viable. The scalar potential in the radial
direction is plotted in \autoref{fig:scalarpot}. Numerically solving
the transcendental equation eq.~\eqref{eq:VDmin} in combination with
eq.~\eqref{eq:tauandt}, we determine the minimum of the scalar
potential at
\begin{equation}
  \taumin{D} \approx 52.71
  \quad \Rightarrow \quad
  \taumin{1} \approx 10.54
  ,\quad
  \taumin{2} \approx 19.33
  ,\quad
  \taumin{3} \approx 22.84
  .
\end{equation}
To verify the above result we also performed a numerical minimization
of the scalar potential eq.~\eqref{eq:scalarpot} in the 3-dimensional
K\"ahler moduli space using Mathematica for the same choice of input
parameters. We again found a metastable de Sitter minimum with
\begin{equation}
  \tau_1^\text{num}\approx 10.54
  ,\quad
  \tau_2^\text{num}\approx 19.33
  ,\quad
  \tau_3^\text{num}\approx 22.84
  .
\end{equation}
The gravitino mass and the cosmological constant at the minimum are given by:
\begin{equation}
  m_{3/2} = e^{K/2}|W| \approx 12 \TeV
  ,\quad
  V\big(\taumin{D}\big) \approx 1.361 \times 10^{-30} m_\text{Pl}^4
  ,
\end{equation}
where we have absorbed the overall factor $e^{{\cal K}/2}$ coming from
the complex structure and the dilaton parts of K\"ahler potential into
the effective parameters $W_0$ and $A$. Note that the vacuum energy,
while huge compared to the actual value, is small in Planck
units. Therefore, much smaller values can be reliably attained by a
further tuning of $W_0$, $A$ and $D$, \emph{without} affecting the
moduli vevs and the gravitino mass as well as all phenomenologically
relevant parameters such as the masses of superpartners. The
feasibility of such tuning by fluxes is explained by the
Bousso-Polchinski mechanism~\cite{Bousso:2000xa}.

Let us now compute the K\"ahler moduli and axion spectrum. The three
canonically normalized moduli eigenstates have masses
\begin{equation}
  m_{\Psi_1}\approx 64\, m_{3/2}\approx 785.6 \TeV
  ,\quad
  m_{\Psi_2}=m_{\Psi_3}
  \approx 1.94\, m_{3/2} \approx 23.7 \TeV
  .
\end{equation}
The heavy eigenstate $\Psi_1$ corresponds mostly to the breathing mode
of the divisor volume $\tau_D$ while the modes $\Psi_2$ and $\Psi_3$
are mostly volume preserving. On the other hand, since the
superpotential contains only one linear combination of the axions,
only one of the axions receives a mass while two of the remaining
eigenstates remain massless.  Indeed, we find that the canonically
normalized axion eigenstates have masses
\begin{equation}
  m_{\Phi_1}\approx 66\, m_{3/2}\approx 809.7 \TeV
  ,\quad
  m_{\Phi_2}=m_{\Phi_3}=0
  .
\end{equation}
The flat directions $\Phi_{2,\,3}$ are the Goldstone bosons arising
from the two shift symmetries that leave the linear combination
$\chi_1+\chi_2+\chi_3$ invariant. It is important to realize, however,
that the superpotential eq.~\eqref{eq:super} must be regarded as the
dominant contribution in a series of terms with other highly
subdominant non-perturbative contributions. These subdominant
contributions will eventually fix the remaining axions, albeit at
vastly suppressed scales. We will comment more on this in
\autoref{phenocosmo}. Finally, after rotating $\tilde{K}_i \equiv
U^{\dag}_{ik}K_{kl}U_{lj}$, $U\in U(3)$, into the axion mass
eigenstate basis the axion decay constants are given by
\begin{equation} 
  f_{\Phi_i}=\sqrt{2\widetilde{K}_i}\,m_\text{Pl}
  \quad\Rightarrow\quad
  f_1\approx 1.7\times 10^{17} \GeV
  ,\quad
  f_2\approx 9.8\times 10^{16} \GeV
  ,\quad
  f_3\approx 9.1\times 10^{16} \GeV
  ,
\end{equation}
which is expected in string compactifications with a high string scale
$M_s \gtrsim M_{GUT}$.

\section{Consistency Condition for the Single Condensate/Instanton Approximation}
\label{consistency}

As mentioned earlier, the superpotential eq.~\eqref{eq:super} must be
regarded as the leading term in a series of contributions. In the
presence of other contributions, it is important to show that the
truncation to the superpotential can be made parametrically
self-consistent. In order to illustrate that, let us assume that the
superpotential contains two non-perturbative terms such that
\begin{equation}
  W = 
  W_0
  -Ae^{-\frac{2\pi}{N}\sum_{i=1}^{\hplus{11}} n_i \tau_i}
  +Be^{-\frac{2\pi}{M}\sum_{i=1}^{\hplus{11}} m^i \tau_i }
  ,
\end{equation}
where we have fixed the axions so that $W$ is real. Substituting the
moduli vevs from eqns.~\eqref{eq:tauitauDprop} and~\eqref{eq:tauDmin}, we
obtain
\begin{equation}
  \label{eq:Wsuppp}
  W
  =
  W_0
  -Ae^{-\frac{2\pi}{N}\tau_D}
  +Be^{-\frac{2\pi}{M}\tau_D\left(\frac 23\sum_{i=1}^{\hplus{11}}
      \frac{m_i a_i}{n_i} \right)}
  .
\end{equation}
Assuming $m_i\sim n_i\sim {\cal O}(1)$, we get $\frac{2}{3}\sum
\frac{m^i a_i}{n_i} \sim {\cal O}(1)$.  Substituting an approximate
expression for the divisor volume $\tau_D\approx \frac
N{2\pi}\ln|2A/3W_0|$ into eq.~\eqref{eq:Wsuppp} we get the following
approximate expression for the superpotential at the minimum:
\begin{equation}
  W 
  \approx 
  W_0
  -\frac{3}{2} W_0
  +B\left|\frac{3W_0}{2A}\right|^{{\frac NM}\times {\cal O}(1)}
  .
\end{equation}
Therefore, when $N/M>{\cal O}(1)$, the extra condensates/instantons
become exponentially suppressed relative to the leading contribution.
So, these contributions do not affect the moduli vevs and only help to fix the 
remaining axions. Note that this is qualitatively different from the case when there 
are many (such as $\hplus{11}$) \emph{comparable} terms in the superpotential. 
A natural setup in which this may arise occurs when the leading
contribution comes from a condensing gauge group with a dual Coxeter
number $N={\cal O}(10)$ while the truncated terms arise from
instantons (that is, $M=1$). The single condensate approximation,
therefore, can be naturally made parametrically self-consistent and
should be quite robust.

\section{Phenomenological Consequences}
\label{phenocosmo}

\subsection{Mediation of Supersymmetry Breaking}

Since the moduli have been stabilized, one can hope to make contact
with particle physics and cosmology, at least in a broad sense. This
requires a specification of the matter and gauge sector - both visible
and hidden. Within the framework considered in the paper, we envision
a four-dimensional matter and gauge sector arising from D3 branes at
singularities or on stacks of D7-branes wrapping four-cycles in the
Calabi-Yau threefold.  From the analysis above, we have found that all
moduli (including all K\"ahler moduli) can be stabilized by
appropriate background fluxes as well as non-perturbative
contributions arising from Euclidean D3-brane instantons or strong
gauge dynamics on a stack of D7-branes, wrapping a single ample
divisor.  For the case of a visible sector arising from D7-branes, one
could imagine stacks of intersecting D7-branes wrapping different
four-cycles, whose volumes are also fixed by the above mechanism,
which may support the visible sector. The zero-mode spectrum of the
visible sector can be made chiral if the Riemann surface that is at
the locus of an intersection between two stacks of D7-branes has a
non-trivial $U(1)$ magnetic flux.  Thus, it is possible to engineer
(semi)-realistic matter and gauge spectra.

What can be said about the issues of supersymmetry breaking and its
mediation to the visible sector? Within the framework above,
supersymmetry is broken by anti D3-branes at the tip of a warped
throat generated in these flux compactifications, and could naturally
be at the TeV scale. By gauge-gravity duality, this is dual to
supersymmetry breaking states in an appropriate quiver gauge
theory~\cite{DeWolfe:2008zy}. Depending on the location of the visible
sector four-cycle relative to that of the anti D3-branes, different
mediation mechanisms could dominate. For example, if the visible
sector resides in the bulk of the Calabi-Yau then the mediation
mechanism is \emph{moduli} (gravity) mediation, suppressed by $1/m_\text{Pl}$
interactions~\cite{Choi:2005ge, Choi:2004sx}. Within such a setup, it
has been argued that the warped throat between the visible and
supersymmetry breaking sectors could give rise to
\emph{sequestering}~\cite{Kachru:2007xp}. It could happen that the
visible sector also resides at the tip of the warped throat, in which
case the dominant mediation mechanism is \emph{gauge}
mediation~\cite{Buican:2007is, Diaconescu:2005pc} arising from
messenger particles stretching between the visible and supersymmetry
breaking sectors. One could also have more complicated setups in which
the visible sector is comprised of a gauge sector on a stack of
D7-branes which extends (partially) in the throat, and a chiral matter
sector which resides in the bulk of the Calabi-Yau. In this case, the
dominant mediation mechanism is \emph{gaugino} mediation, from
exchange of gauginos extending in the throat~\cite{Benini:2009ff,
  McGuirk:2009am}. Thus, there can be a wealth of possibilities for
low-energy particle physics. The situation in Type IIB
compactifications is quite different from that in fluxless $G_2$
compactifications of $M$-theory where warping is not expected to be
present. In $M$-theory compactifications to four dimensions, matter
and gauge sectors live on three-cycles, and two three-cycles
generically do not intersect in seven dimensions. Hence supersymmetry
breaking in the hidden sector is mediated to the visible sector by
moduli fields, giving rise to \emph{gravity}
mediation~\cite{Acharya:2008hi, Acharya:2007rc}.

\subsection{Dynamical Solution to the Strong CP-Problem}

We now point out some consequences of the above moduli stabilization
mechanism, which are different from the other mechanisms within Type
IIB string theory, but have features similar to that in $M$-theory. As
found in \autoref{modulivevs}, the K\"ahler modulus $\tau_D$ and its
axion partner receive a mass of ${\cal O}(10)\times m_{3/2}$, while
the remaining K\"ahler moduli receive masses of ${\cal
  O}(m_{3/2})$. The remaining axions turn out to be massless at this
level, but will be eventually fixed by possible subdominant contributions to
the superpotential. Note that this is quite different from the
standard scheme in which one has $\hplus{11}$ \emph{comparable}
non-perturbative terms in the superpotential, each depending on a
different K\"ahler modulus. In that case, as stated earlier all axions,
which are imaginary parts of the K\"ahler moduli, are stabilized at
${\cal O}(m_{3/2})$.

This property of the axion spectrum has a crucially important
consequence for a dynamical solution to the strong CP-problem. It has
been a long-cherished dream in string phenomenology to use one of the
numerous axions arising in a string compactification to be the QCD
axion. One of the crucial requirements for the axion to be a QCD axion
is that the dominant contribution to its potential must arise almost
entirely from QCD instanton effects, with a mass given by
$m_{a,\text{QCD}}\sim {\Lambda_{QCD}^2}\big/f_{a,\text{QCD}} \ll
m_{3/2}$. However, within the standard scheme mentioned above the
axions are already stabilized at ${\cal O}(m_{3/2})$, so none of them
can serve as QCD axions. The moduli fixing mechanism described in this
paper, on the other hand, only fixes one combination of axions at
${\cal O}(10)\times m_{3/2}$, so this serves as an excellent starting
point for a solution to the strong CP-problem. It turns out that the
remaining axions are stabilized at exponentially suppressed scales
relative to $m_{3/2}$, which is especially true when the subdominant
non-perturbative contributions come from instantons as opposed to
gaugino condensates, as should be evident from
\autoref{consistency}. Furthermore, recall that there is a
field-dependent prefactor for the non-perturbative contribution for
the chiral visible sector. Since these fields must have vanishing vevs
for phenomenological reasons, the instanton contribution from the
visible sector four-cycle vanishes in the vacuum. Thus, the linear
combination of axions that corresponds to $\theta_{QCD}$ receives its
mass entirely from the QCD instanton effects, thereby solving the
strong CP-problem.

Another very interesting consequence of the mechanism is the presence
of a \emph{multitude} of very light axions, with masses distributed
roughly linearly on a logarithmic scale. Thus, the framework
dynamically realizes the ``String Axiverse'' scenario discussed
in~\cite{Arvanitaki:2009fg}, with many of their observable
signatures. Note that the above features also hold for fluxless $G_2$
compactifications of $M$-theory since the moduli and axion fixing
mechanism is very similar to that discussed in this paper. A full
analysis of the solution to the strong CP-problem including observable
consequences vis-a-vis the above mechanism has been carried out
in~\cite{axions}.

\section{Conclusions}\label{conclude}

In this work we have proposed an elegant and robust mechanism to
stabilize all K\"ahler moduli in Type IIB string compactifications on
Calabi-Yau orientifolds with D3/D7-branes, which is motivated from the
analysis of moduli stabilization in fluxless $G_2$ compactifications of
$M$-theory. This can be achieved with just one non-perturbative
contribution to the superpotential, arising either from D3-instantons
or from strong gauge dynamics on D7-branes, wrapping an ample divisor
with $\chi_+-\chi_-=1$. This scheme also naturally provides a
dynamical solution to the strong CP-problem within string theory.

In order to break supersymmetry and obtain a de Sitter vacuum, we have
followed the KKLT proposal and included explicit supersymmetry
breaking terms in the scalar potential due to anti D3-branes. It has
been argued recently that such a procedure may not be consistent from
a microscopic point of view~\cite{Bena:2009xk}. It is, therefore, important to study
mechanisms in which supersymmetry is broken spontaneously and a de
Sitter vacuum is obtained by adding additional $F$-term or $D$-term
contributions to the potential. While this will not change the
qualitative picture of moduli and axion stabilization and the
resulting solution to the strong CP-problem outlined in the paper, it
could have important effects for particle physics. This is being
worked out in a companion paper~\cite{PartTwo}.

\acknowledgments 

We would like to thank Bobby Acharya, Ralph Blumenhagen, Ben Dundee,
Joe Marsano, and Timo Weigand for useful discussions. K.B. and
S.R. are supported by DOE grant DOE/ER/01545-883. P.K. is supported in
part by the US Department of Energy under Contract DE-AC02-05CH11231
and in part by the US National Science Foundation Grant PHY-04-57315.

\bibliography{PartOne}

\end{document}